# Neural-FacTOR: Neural Representation Learning for Website Fingerprinting Attack over TOR Anonymity

Haili Sun
School of Cyber Science and Engineering
Huazhong University of Science and Technology
Wuhan, China
hailisun@hust.edu.cn

Yan Huang
School of Artificial Intelligence and Automation
Huazhong University of Science and Technology
Wuhan, China
platanus@hust.edu.cn

Lansheng Han*
School of Cyber Science and Engineering
Huazhong University of Science and Technology
Wuhan, China
hanlansheng@hust.edu.cn

Xiang Long
School of Cyber Science and Engineering
Huazhong University of Science and Technology
Wuhan, China
longxiang@hust.edu.cn

Hongle Liu
School of Cyber Science and Engineering
Huazhong University of Science and Technology
Wuhan, China
Lhl753159@gmail.com

Chunjie Zhou
School of Artificial Intelligence and Automation
Huazhong University of Science and Technology
Wuhan, China
cjiezhou@hust.edu.cn

*Abstract*—TOR (The Onion Router) network is a widely used open source anonymous communication tool, the abuse of TOR makes it difficult to monitor the proliferation of online crimes such as to access criminal websites. Most existing approches for TOR network de-anonymization heavily rely on manually extracted features resulting in time consuming and poor performance. To tackle the shortcomings, this paper proposes a neural representation learning approach to recognize website fingerprint based on classification algorithm. We constructed a new website fingerprinting attack model based on convolutional neural network (CNN) with dilation and causal convolution, which can improve the perception field of CNN as well as capture the sequential characteristic of input data. Experiments on three mainstream public datasets show that the proposed model is robust and effective for the website fingerprint classification and improves the accuracy by 12.21% compared with the state-of-the-art methods.

*Keywords—security protection, anonymity attack, website fingerprinting, TOR, neural representation learning*

## I. INTRODUCTION

With the increasing popularity of network and 5G communication technology, more and more attention has been paid to the research of network security protection. To protect personal privacy from censorship and eavesdropping, network users tend to communicate via anonymous techniques, such as the widely used privacy-aware tool TOR[1]. As one of the most widespread anonymous communication technologies, TOR protects user's privacy through separating their personal information from their visiting behavior. It is composed of volunteer nodes which relay encrypted traffic between the web server and the client to ensure no node take the right to know both the destination and origin[2].

However, an attacker can still speculate which specific website a user has browsed by analyzing the TOR's side-channel traffic[3]. A local passive attacker can infer the user's destination without decryption by observing and analyzing the packet sequences generated between client and entry node of the TOR network[4], which is called as website fingerprinting (WF) attack.

WF attack acts as a classification technique which is used to eavesdrop on the online behavior (i.e. browsing a website on an encrypted network) of the user by analyzing special traffic. The attacker gathers the client's traffic traces as packets sequences. By classifying these sequences, the inspector estimates which website the user is visiting. For example, a user may be visiting a video site if he sends and receives many packets in a short time and the interval is very large. On the contrary, he may be viewing a web page or checking an email when few packets are transmitted with a small interval.

WF attack is feasible and effective for TOR de-anonymization which has been proven in many researches[5], [6]. Most of the proposed attack approaches[7][8] adopt machine learning (ML) techniques to recognize association pattern between personal characteristics and website fingerprint. They extract manually selected features of the packet sequences and then classify them with ML algorithms. However, manually selected features are commonly resort to researchers' own experience, intuition or expert knowledge and are usually expensive[9]. Moreover, Juares and Dye[10], [11] point out that the accuracy lies to a large extent on the extracted features rather than a particular machine learning algorithm. Thus, these attacks may abort if defense attempt to conceal these specific features. To automatically extract features, Rimmer, Preuveneers, Juarez, Van Goethem and Joosen[12] propose to employ Long Short-Term Memory (LSTM), Stacked Denoising Autoencoder (SDAE) and Convolutional Neural Network (CNN), demonstrating the availability and advantage of adopting deep learning (DL) algorithm to WF attacks. After that, many researchers[13]-[16] have dedicated to achieve higher accuracy by improving these DL models.

Nonetheless, how to utilize the diverse features hidden in the TOR traffic is still an open problem. The significant sequential features and statistical features (e.g. time-order characteristics and total packet size) have not been fully explored for a long time. Hayes & Danezis[9] pointed out that the packet interval time merely raise the attack accuracy by a small degree.

Consequently, in this paper, we propose an effective neural representation learning model named Neural-FacTOR for WF attack over TOR anonymity, which utilize not only statistical features but also direction information of the packet sequence. Causal and dilated convolution are adopted to further expand the receptive field of neuron and learn sequential feature of direction sequences. Experimental results on public datasets including Rimmer7, MutiLayer20 and Wang35 show that our model can benefit greatly from statistical features and temporal characters. Compared with the state-of-the-art attack methods, the proposed attack approach is confirmed to perform better with less time cost.The contributions of our work can be summarized as follows:

- Propose an effective website fingerprinting attack based on CNN which can automatically extract features from raw inputs avoiding the bottleneck of manually portraying data feature set. Furthermore, we add some statistical features of the packets while encoding for better discrimination power, to improve the final classification accuracy.

- To reduce the overhead and extract temporal characteristics between raw packets, we adopt dilated and causal convolutions which are beneficial to decrease training parameters and extract sequential features from raw packets boosting the convergence speed and improving performance of model.

- Demonstrate the effectiveness of dilated convolution for the website fingerprinting attack task, i.e. promoting the convergence speed as well as achieving better classification performance with fewer parameters than the state-of-the-art models.

---
* The corresponding author

The rest of the paper are organized as follows. In Section Ⅱ, we present the background information and related works. Then, we discuss the motivation and insights of our work and the proposed website fingerprinting attack model base on CNN in detail in section Ⅲ. Section Ⅳ and Ⅴ evaluate and analyze the proposed model, respectively. Finally, Section Ⅵ concludes the paper.

## II. RELATED WORK

In this section, we explain the threat framework for WF attack and review the existing WF attack models on TOR.

### A. Threat Framework

Since WF attacks can identify the website a user is visiting, it is commonly used to eavesdrop on the user's online behavior on a encrypted network, such as Tor, which is a prevalent anonymous communication network allowing the user to conceal the source and destination traffic by encrypting communication packets and building virtual circuit. To carry out a WF attack, an adversary eavesdrops packets from monitored websites, learns characteristic from collected packets and trains ML techniques to distinguish whether the user is browsing a specific website among the monitored ones.

As shown in Fig. 1, the adversary acts as a local passive inspector, who eavesdrops the traffic between the victim and the TOR network. A local attacker can have access to the entry node which the client selected to connect to the TOR (such as Adversary 2), or the link between the entry node and the victim (such as Adversary 1 in Fig. 1). The user only collects the packets transmitted during user's browsing behavior without modifying the traffic flow. Thus, this attack mode is hard to be detected. Besides, in real scenario, a WF attacker may be an Autonomous System (AS), an Internet Service Provider (ISP), or a router, etc. To simplify this issue, we assume that the user only browse one web page each time without any other behaviors, like picture uploading.

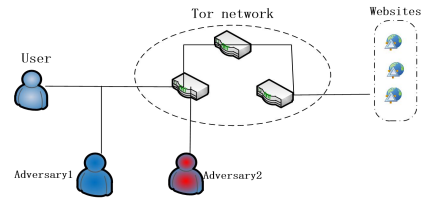

Fig. 1. The threat framework of website fingerprinting attack

### B. Existing WF attacks

KNN-WANG[17] constructs a 4226-sized feature set from side-channel information, and adopts a k-NN (k-Nearest Neighbor) classifier to achieve an accuracy of more than 90% on 100 websites. However, it has the following shortcomings: 1) rely on a large feature set, which is time-consuming with high computation complexity; 2) treat all features equally with no highlight on some special characteristics which are significant, e.g. total transmission time. Besides, CUMUL[18] introduces a custom RBF core for SVM algorithm to construct WF attack model. This method indeed reduces the feature set, however, data pre-processing truncation is too cumbersome and each sample needs an "add-up" calculation, which is too costly for implementation.

K-FP-Hayes[9] uses random forest to calculate features' importance which have not appeared in existing works. Moreover, they propose to add time series characteristics into feature sets and is shown to be effective comparing to CUMUL algorithm with higher robustness. However, this method needs to train model twice which renders very expensive calculation and time cost. Once the content of the target site changed, the cost of retraining model can be enormous. Besides, it adopts k-NN algorithm for final classification which further explains the selection and construction of feature set is far more important than the options used in traditional website fingerprinting classification models. Reference [2] proposes a new method for WF attack named 2ch-TCN, which integrates both packet direction sequence and time information, but still gains an unsatisfied accuracy. Moreover, its training process is time consuming due to the too many layers stacked and residual connection.

To distinguish with the related works, our work has the following differences:

- Adopt causal convolution to capture the time-order characteristics of packet sequence, which can better handle temporal dependencies and have not been considered in existing works to the best of our knowledge;

- Expand the receptive field of neurons by employing dilated convolution, which can promote the convergence speed with fewer training parameters and no performance degradation;

- Exploit some important statistical features including the size and number of requested packets for input encoding and is shown to be effective for improving the accuracy of anonymity attack.

## III. DESIGN OVERVIEW

In this section we introduce the motivation of our work at first and then describe the details of the proposed attack model. This paper models WF attack task as a special sequence classification task. The model needs to automatically learn and fit the high-level features of the packet sequence. Then, it predicts the category of an unknown sequence pair indicating a successful attack.

### A. Motivation and insight

The widespread use of deep learning is largely due to its self-learning ability for features. It can learn the characteristics and associations within data from raw input, which is difficult for human to describe and understand. Further, the practical application results show that the features learned by deep learning algorithm tend to perform better and are more robustness than manually extracted ones.

Inspired by [19] which utilizes CNN network to analyze encrypted video sequences, we adopt CNN network to learn the WF from encrypted packets sequence to analyze the user's browsing behavior. Therefore, in this paper, we propose a new website fingerprinting classification model based on dilated convolution and causal convolution.

### B. Our Neural-FacTOR model

#### 1) Dilated and causal convolution

Dilated convolution and causal convolution are two variants of CNN which are suitable for the TOR network de-anonymization task based on website fingerprinting.

*a) Dilated convolution:* This classic variant[20] gains a large receptive field and increases kernel size by expanding the alignment of the kernel weights with dilation parameter[21].

*b) Causal convolution:* The ordering feature of packet sequence context can be extracted by causal convolution without recurrent connection. In this paper, we use causal convolution to handle the order of packet sequence.

#### 2) Details of Neural-FacTOR

CNN is a typical and efficient encoding model for many computing tasks, which can automatically extract sequence features from original input[22] and build representation vector for specific task. Nevertheless, there also exists two issues unsolved for higher accuracy. One is the limited receptive filed of a neural unit in CNN model, and another is that the features are insufficient which are learned only from the direction sequence (of packet/cell).

To address the first issue, we employ dilated convolution that enable an exponentially large receptive filed[20]. For the second issue, we utilize causal convolution to extract the order characteristics from packet sequence. Inspired by [9], we also select some important side-channel information (e.g. total packet size) as input data, which is beneficial to further improve the accuracy.

Then, we utilize fully-connected layer to integrate the statistical features and the output of packet sequences representation model. In addition, a softmax function is used to map the integrated representation to the corresponding class labels. Furthermore, we also adopt dropout mechanism[23] to avoid overfitting. Benefit from the outstanding sequence feature extracting capability of dilated convolution and causal convolution, our model could perform better than existing methods.

The architecture of the proposed Neural-FacTOR[2] model is depicted in Figure 2. As is shown, our model integrates the statistical features and direction sequence features of input data. To fit large amount dataset as much as possible, we use 12 convolution layers, and each layer is setted with 4 filters and 3 size of kernel. Additionally, for dilated convolution layers, to avoid the dilated gridding effect[26], we adopt hybrid dilated convolution strategy to design the dilation rate. Thus, a dilation rate list: {1, 2, 4, 8} is provided and the dilation rates of those layers would be setted following the list that can lead to gain richer information from a wider perspective. For model training, the categorical cross-entropy is exploited as loss function and Adam[24] is adopted as optimizer with learning rate 0.001 which is suitable for learning the characteristics of TOR traffics shown by [2][12].

## IV. EXPERIMENTAL SETUP

We present the steps for data processing and datasets used for experiments in this section.

### A. Data preprocessing

As shown in Figure 3, there are three layers (Cell layer, TLS layer, TCP layer) in TOR network for data transmission. The application data is first divided into several fixed-length cells in Cell layer as they are the smallest units in TOR. After that, for transmission security, the encapsulated cells will use the TLS protocol for secure transmission. A TLS record may contains one or more cell, or it may also be zero (the TLS record at this time is a non-data type record). Finally, each TLS record is encapsulated into one or more TCP data packet in TCP layer, the size of which is limited by the maximum segment size.

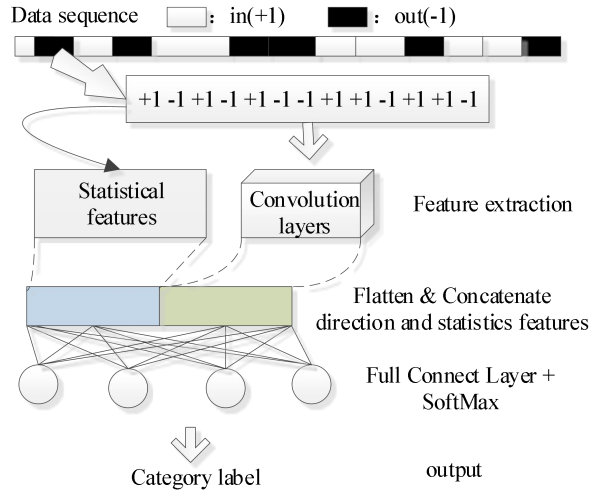

Fig. 2. Architecture of the Neural-FacTOR model

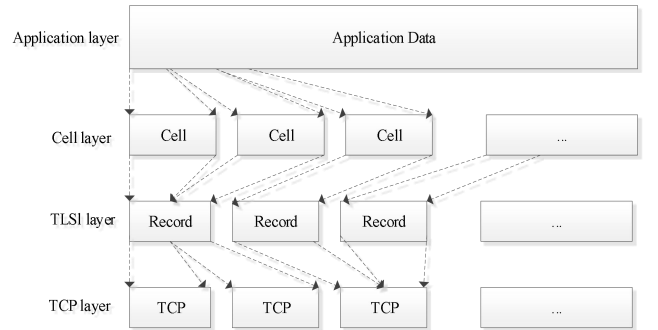

Fig. 3. Layers of Data Transport in TOR

---

[2] Our code can be found here: https://github.com/amykalar/Neural-FacTOR

*B. Dataset*

We evaluate and compare the proposed attack model with several state-of-the-art methods on three different datasets: Rimmer et al.'s dataset[12] (referred to as Rimmer7), MultiLayer20[2] dataset and Wang et al.'s k-NN[17] dataset (referred to as Wang35). We introduce the details of these datasets in this section.

*1) Rimmer7:* The Rimmer7 dataset contains 900 websites, each with 2500 valid TOR network access records. These 900 websites are selected from the homepage of 1,200 of the most popular websites according to Alexa. The subset was collected through a distributed setup with 240 worker threads of 15 virtual machines. The records in *Rimmer7* are from cell layer, only containing the direction information of cells. For facilitating comparison with Wang35, we select 100 websites with 2,500 instances each.

*2) MultiLayer20:* This dataset is consist of two subsets of monitored websites: 1,400 traces from each of 850 popular TOR hidden services and 500 top Alexa monitored websites, each site with 2,300 traces. Each trace include both time and direction information of packets. The dataset is collected by using distributed virtual machines whose locations ranging from Japan, France, Singapore, Australia, the United States and the Netherlands. Besides, MultiLayer20 contains three extraction layers (Cell, TLS and TCP layer) traffic data.

*3) Wang35:* The *Wang35* dataset has long been the primary dataset used to evaluate the effectiveness of website fingerprinting attack. It consists of 100 sensitive pages each with 90 records for a closed-world evaluation and 5,000 non-monitored pages each with 1 records for open-world experiment. Those 100 monitored pages were compiled from some preselected blocked web pages from Saudi Arabia, China, and the UK while the 5000 non-monitored pages were selected from Alexa's top 10,000. Traces in *Wang35* are cell sequences, also containing both the direction and timestamp information of each cell. The traces were collected in batches, each circuit is tried to use as long as possible and the same circuit is maintained in one batch. Additionally, they collected all the traces by using one client with a fixed IP.

## V. EVALUATION AND DISCUSSION

In this section, we evaluate our WF attack Neural-FacTOR on Rimmer7 and Wang35 datasets with both *closed-world* and *open-world* experiments. We also compare the results of our attack model with state-of-the-art attacks and the comparison results show that our model is excellent in classification accuracy with relatively small overhead. Besides, we reevaluate our model and previous attacks on the MultiLayer20 dataset.

From the comparison of attacks in three traffic data layers, we draw two conclusions:

- The raw traffic data indeed contains more information (e.g. the real size of each packet) than normalized data that each packet size is normalized to 1;
- The TCP layer is more suitable for classifier than the other two layers.

*A. Comparison with existing Attacks*

In this section, five state-of-the-art models are chosen for comparing with our proposed attack model, including Var-CNN[14], CUMUL[18], 2ch-TCN[2], CNN and SDAE[12] (denoted as CNN-Rimmer and SDAE-Rimmer respectively to avoiding confusion with common neural network models). All these baseline attack models are most typical methods based on deep neural network (except CUMUL) with SOTA (state-of-the-art) accuracy, and CUMUL model is a superior attack model proven by Rimmer et al.[12].

Accuracy is used to measure the performance of these models. For Rimmer7 dataset, we choose 375 packets of each website for testing and validation respectively and 1,750 packets for training. For Wang35 dataset, 30 instances of each website are chosen for testing and another 60 instances for training and validation.

As shown in Table I, obviously, our model outperforms the state-of-the-art models on the two datasets. Besides, the accuracy of all attacks on the Rimmer7 dataset is higher than that on the Wang35 dataset, as the number of trace for each website increases in the former than the latter. This phenomenon shows these classifiers are feasible with larger scale packet sequence.

Our model, Var-CNN-dir, CUMUL and 2ch-TCN can still perform well although with fewer packet per site. On the contrary, CNN-Rimmer and SDAE-Rimmer do not perform well. When the number of packets grows, the two attacks gain significant improvement, which are consistent with the experimental results in [5]. Based on this, we can infer that the data scale plays a strong impact on Rimmer's attacks.

*B. Experimental Results on MultiLayer20 dataset*

In this subsection, five most promising deep learning based attacks: TCN-dir, 2ch-TCN, Var-CNN-dir, Rimmer-SDAE and our Neural-FacTOR are evaluated on the MultiLayer20 dataset, which consists of traffic data in three extraction layers: the TOR cell, TLS and TCP layer. In order to get better performance, we filter the site whose instances less than 100 and the instances with length less than 500. Finally, we obtain 323 website instances in cell layer, 114 in TLS and 193 in TCP layer, each site contains more than 100 instances. For experiment, we use 70 instances of each site for training and 30 instances for testing in different layers. Note that the public MultiLayer20 dataset has been desensitized and confused.

Table II shows the experimental results on MultiLayer20 dataset. As is shown, the classification accuracy of most attacks has decreased. This may be caused by the mechanism that for each visit a new circuit is established during data collection process. Comparing with the baseline models, our model Neural-FacTOR still performs better, which achieves the highest accuracy of 93.80% in TCP layer. Furthermore, we obtain an accuracy improvement of up to 12.21% than the state-of-the-art classifier in this layer. In TLS layer, the average accuracy improvement is up to 11.32%.

TABLE II. RESULTS (%) OF ATTACKS ON MULTILAYER20 IN THREE LAYERS

| Attacks | TCP[c] | TCP[d] | TLS[c] | TLS[d] | Wang's Cell[d] | Real Cell[d] |
|---|---|---|---|---|---|---|
| SDAE-Rimmer | 47.25 | -- | 48.6 | -- | 70.66 | 40.35 |
| Var-CNN-dir | 65.84 | -- | 67.98 | -- | 77.59 | **64.92** |
| TCN-dir | 66.77 | 72.85 | 51.58 | 50.91 | 70.30 | 47.75 |
| 2ch-TCN | 81.59 | 77.69 | 55.26 | 55 | -- | 42.19 |
| Neural-FacTOR | **93.80** | **84.04** | **79.30** | **63.83** | **88.91** | 61.02 |
| Improvement | **12.21** | **6.35** | **11.32** | **8.83** | **11.32** | -- |

[c.] employed the raw packets sequence
[d.] employed the converted +1, -1 sequence

TABLE I. ACCURACY (%) OF THE SIX ATTACKS ON WANG35 AND RIMMER7 DATASET

| Attacks | Wang35 dataset | Rimmer7 dataset |
|---|---|---|
| CUMUL | 91.38[a] | 95.43[b] |
| CNN-Rimmer | 71.43 | 91.23 |
| SDAE-Rimmer | 87.78 | 95.50 |
| Var-CNN-dir | 93.20 | 97.01 |
| Var-CNN | 93.33 | - |
| TCN-dir | 90.60 | 92.15 |
| 2ch-TCN | 91.73 | - |
| Neural-FacTOR | 93.33 | 97.25 |

[a.] Based on experiment results from [18]
[b.] Based on experiment results from [12]

This enormous improvement shows that Neural-FacTOR can better learn the features of direction sequence. As the direction sequence itself is chronological, it is enough for us to learn packet representation in both sequential and direction aspects.

When compared with 2ch-TCN[2] which uses 16 layer stacks with residual block, Neural-FacTOR (with 12 layer convolution) is apparently more excellent in classification accuracy with less overhead, as shown in Table II and Table III. Additionally, as shown in Fig. 4, Neural-FacTOR converges steadily during training process. On the contrary, the 2ch-TCN model overfits after 20th epoch. Furthermore, the total parameters of the former is far less than the latter one, as shown in Table IV. This implies that Neural-FacTOR is more robust and steady with fewer parameter than 2ch-TCN.

In addition, from Table II, we also can see that the TCP layer data yield the best result when comparing the results of these four models using different layer data. This shows the TCP layer can provide more deterministic information for the classifiers.

Note that we employ the raw packets sequence in TCP and TLS layer rather than convert them to +1, -1 sequence (+1 indicates 'in' packet and -1 indicates 'out'). And as shown in Fig. 6, the comparison results also show that the raw packets sequence contains more information which make a significant contribution for classification accuracy.

### C. Effectiveness of Dilated convolution

In order to improve the perception of neurons in CNN model, we add a dilated convolution to the network, so that the model can learn the connection better without treating the input data as independent points. Meanwhile, to validate the effectiveness of the dilated convolution, we also trained another eight-layer model for comparison.

Figure 5 illustrates the performance of the eight layer and the twelve layer CNN network in the closed world experiment respectively. The experimental results apparently show that the performance of the two models with dilated convolution is undoubtedly better comparing with the other ones. In addition, the accuracy of the eight layers network model with dilated convolution is higher than the twelve layers network model without dilated convolution. This phenomenon indicates the effectiveness of the dilated convolution for WF attack task.

TABLE III. TRAIN TIME COST (S) OF EACH MODEL IN DIFFERENT LAYERS

| Attacks | TCP[c] | TCP[d] | TLS[c] | TLS[d] | Cell[d] |
|---|---|---|---|---|---|
| TCN-dir | 3056 | 3059 | 1823 | 1838 | 5092 |
| 2ch-TCN | 3964 | 3081 | 1820 | 1844 | 5121 |
| Neural-FacTOR | 2357 | 1209 | 1407 | 731 | 4728 |

TABLE IV. PARAMETER NUMBERS OF THE THREE OPTIMAL MODELS

| Attacks | Total | Trainable | Non-trainable |
|---|---|---|---|
| 2ch-TCN | 32,307,801 | 32,305,753 | 2,048 |
| Var-CNN-dir | 4,537,266 | 4,525,426 | 11,840 |
| Neural-FacTOR | 1,935,021 | 1,930,573 | 4,448 |

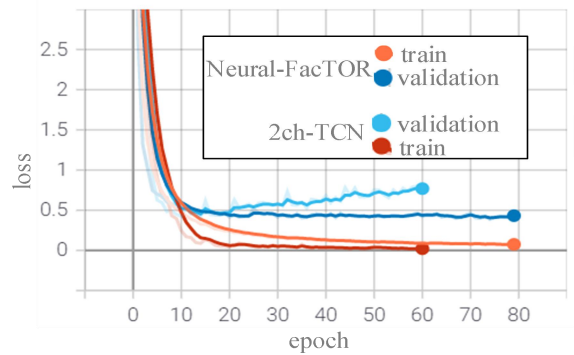

Fig. 4. The loss curve on MultiLayer20 dataset

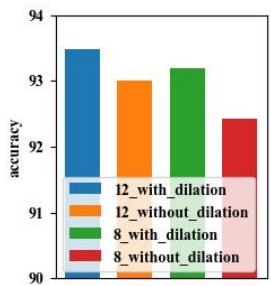 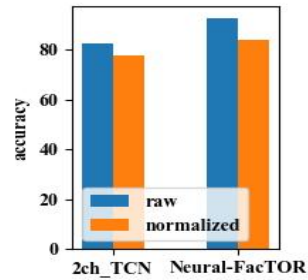

Fig. 5. The effectiveness dilated convolution

Fig. 6. The comparison result in TCP layer

## VI. Conclusion and Future Work

In this work, we propose a novel website fingerprinting attack model, named Neural-FacTOR, which automatically extracts features from packet sequence. Experimental results on three different datasets show that our model can benefit greatly by using causal convolution and dilated convolution. Comparing with the state-of-the-art attacks, our attack model is demonstrated to perform better in both accuracy and convergence speed with fewer parameters. By evaluating our attack on the Multilayer20 dataset in three layers, we observe that the data from the TCP layer is the most divisible among the three layers. Another nice finding is that the raw data (in TCP or TLS layers) contain certain more information than normalized data.

It is promising to employ deep neural representation methods for WF attacks. As WF attack technology is widely used to analyze TOR packet sequences, more and more defense measures have been proposed, such as Walkie-Talkie, DynaFlow[25]. These technologies may interfere with the recognition accuracy of existing WF attack models. Due to the limitation of condition, this article does not test the performance of our WF attack model against these defense methods. In future, we will evaluate the interference of these defense methods on the WF attack model in actual environment.

## Acknowledgments

This paper is supported by National Natural Science Foundation of China: No.62127808, No.62172176, No.62072200.